\pdfoutput=1
\documentclass[%
 reprint, 
nofootinbib,
 amsmath,amssymb,
 aps, physrev,
floatfix,
]{revtex4-2}

\usepackage{orcidlink}
\usepackage{graphicx}
\usepackage{multirow}
\usepackage[caption=false]{subfig}
\usepackage{makecell}

\begin{document}

\preprint{PRAB/XX}

\title{\textbf{Reinforcement Learning applied to Optimization of LHC beams in the CERN Proton Synchrotron} 
}%

\author{\orcidlink{0000-0002-0498-9798}Joel Wulff}
 \email{Contact author: joel.wulff@cern.ch}
\author{\orcidlink{0000-0001-6377-2823}Alexandre Lasheen}%
 
\affiliation{%
  \href{https://ror.org/01ggx4157}{CERN}
}%
\date{\today}

\begin{abstract}
   The longitudinal triple splitting in the CERN Proton Synchrotron (PS) is a key rf manipulation defining the 25 ns bunch spacing delivered to the Large Hadron Collider (LHC). We present an automated optimization of this manipulation based on machine learning. Successive manipulations with rf systems at multiple harmonics of the revolution frequency are performed in the PS. Each bunch injected from the PS Booster (PSB) is split into twelve bunches with ideally identical longitudinal beam parameters. Precise rf voltage and phase settings are required to minimize bunch-by-bunch variations in intensity, longitudinal emittance, and bunch shape. Our setup combines two distinct parts: a convolutional neural network providing an initial phase correction from the evolution of longitudinal bunch profiles during the splitting process, and two sequential Soft-Actor-Critic (SAC) reinforcement-learning agents that refine cavity phases and voltages. The models are trained on data from Beam Longitudinal Dynamics (BLonD) tracking simulations augmented by simulated uncertainties, including noise evaluated from measurements, to enable training and robust transfer to the machine. First tests in 2022 reached target splitting quality in fewer than ten optimization steps on average, matching or outperforming manual adjustments. This led to operational deployment of an on-demand version, followed by a fully autonomous controller in March 2025. This controller has been available to operations since, representing one of the first reinforcement-learning-based systems for beam quality optimization deployed in the CERN injector complex.
\end{abstract}

\maketitle


\section{Introduction}

The longitudinal beam parameters for the LHC demands trains of multiple bunches with target longitudinal emittance of $\varepsilon_l=0.35$~eVs, spaced by 25~ns at PS extraction. Achieving these conditions requires exploitation of the full range of rf systems which span a wide range of revolution frequency harmonics (7 to 21, 42, 84) and enable complex longitudinal manipulations such as batch compression, bunch rotation, and longitudinal bunch splittings~\cite{bunch_mergin_and_splitting, Garoby:rf_gymnastics}.

Many of these manipulations still rely on manual tuning for performance optimization, introducing variability and limiting the reproducibility of beam quality. Combined with varying beam conditions and hardware drifts, this motivates the need for online, automated, and robust tuning procedures.

A specific requirement arises from the constraints on bunch-by-bunch intensity and length at PS extraction, which must remain within $\pm$3\% RMS (about $\pm$10\% total spread)~\cite{Buffat:2802720}. These variations are caused by the intensity spread of the bunches from the previous accelerator, the PS Booster (PSB), and the bunch splittings in the PS. Two types of longitudinal splittings are employed. Each bunch first gets split into three, then passing through two sequential double splittings, giving the total splitting factor of $3\cdot2\cdot2=12$.

For a double splitting, two rf systems with a harmonic ratio of two are pulsed simultaneously. The bunch is initially held by the lower-harmonic system, while the voltage of the higher harmonic is zero. The unstable phase of the higher-harmonic system is located at the initial bunch center. As the voltage of the second cavity is slowly increased, the voltage of the first cavity is decreased. This lengthens the bunch before progressively splitting it in two. Thanks to a linear correlation between the relative phase error of the active rf systems and the final bunch-intensity imbalance, the manipulation is relatively straight-forward to optimize, requiring only tuning of a single phase setting with access to an informative observable.

The triple splitting, a manipulation unique to the PS, is considerably more complex. It requires three simultaneously active rf systems at harmonics ratios 7, 14 and 21. As for the double splitting, the bunch is initially captured by the lowest-harmonic system, and is subsequently transferred to the higher harmonics through programmed voltage ramps. The second harmonic ($h=14$) is pulsed in counter-phase with respect to the first, pushing particles toward the outer buckets while the third harmonic ($h=21$) provides the final capture (see Fig. \ref{fig:triple}). The phase and voltage parameters are all coupled, and influence not only the relative intensities but also the final bunch distributions and longitudinal emittances produced by the manipulation. As a result, manual tuning of the triple splitting is in practice time consuming, difficult to standardize, and increasingly inadequate as beam-quality demands tighten.

Previous attempts to automate the optimization of these manipulations using conventional numerical methods have achieved limited success. Although suitable objective functions could be designed for the double splitting, the triple splitting proved substantially more difficult due to its high dimensionality and non-linear parameter couplings. Moreover, the optimization algorithms explored previously did not converge quickly enough for operational use, motivating the search for alternative approaches capable of faster and more robust convergence to good beam quality~\cite{lasheen:ipac2021-wepab244}.

Machine-learning methods offer a promising path forward. They allow transferring the exploration burden to an offline training phase, where models learn a mapping from states to actions from a representative dataset (or interactions), thereby avoiding costly online exploration during deployment. In contrast to numerical optimization, this enables rapid convergence at inference time. Supervised neural networks can map observations to beam or system parameters when sufficient labeled data are available~\cite{lasheen:ipac2021-wepab244, slac_SL_phase_space, incorp_cnn_img_based_diagnostics, kaiser:ipac2023-thpl019, virtual_diag_maxiv}, but their generalization is constrained by the quality and coverage of that labeled dataset. Reinforcement Learning (RL) offers a complementary approach: instead of relying on labels, RL learns control policies through interactions with an environment, guided by a reward function. This makes RL particularly appealing for accelerator control tasks, where direct and fast optimization of system behavior under varying beam/equipment conditions is often required.

Here, we developed an automated method for optimizing the longitudinal rf triple splitting in the PS, using a combination of supervised and reinforcement learning to achieve fast and reliable tuning across a wide range of beam conditions. A quantitative measure of the splitting quality is introduced and used both to define optimization targets and to guide reward shaping during RL training. Although supervised and RL-based approaches were initially evaluated independently, robust performance was achieved only when they were combined.

The final implementation to optimize the triple splitting consists of three machine-learning models applied sequentially: a supervised Convolutional Neural Network (CNN) providing an initial coarse phase correction, followed by two RL agents, one optimizing the relative phases of the active RF systems, and the other adjusting the voltage of the intermediate-harmonic system. All models are trained exclusively on domain-randomized simulated data (generated using the Beam Longitudinal Dynamics tracking code BLonD~\cite{blond}) and transferred directly to the accelerator without further fine-tuning. The approach has been validated on various LHC beam schemes, from 72- to 36-bunch trains at extraction and at different intensities, demonstrating fast and reliable optimization. Since early 2023, this system has been operationally deployed to optimize both dedicated test beams and standard LHC beams, constituting one of the first RL-based beam-quality optimization tools to be routinely used at CERN. Since early 2025, it has been available as a fully autonomous controller, capable of activating and maintaining optimal splitting quality without operator intervention.

Section II describes the triple splitting manipulation in the PS, introduces the quality metric applied to evaluate splitting performance, and shows how the optimization problem can be decoupled into separate phase and voltage sub-tasks. Section III details the simulation and machine learning framework, including the generation of the training dataset, the domain randomization techniques applied to bridge the gap between simulation and measurement, the architecture and training of the supervised feature extractor CNN, and the design and training of the two RL agents. Section IV describes the operational deployment of the combined scheme, first as an on-demand optimizer and subsequently as a fully autonomous controller. Section V presents results from a controlled stress test and from three months of operational experience with LHC beams. Finally, Sections VI and VII discuss the broader implications of the approach, its current limitations, and possible directions for future development.

\section{The triple splitting in the PS}
\label{sec:background}

\begin{figure}[!htb]
    \centering
    \includegraphics[width=0.95\linewidth]{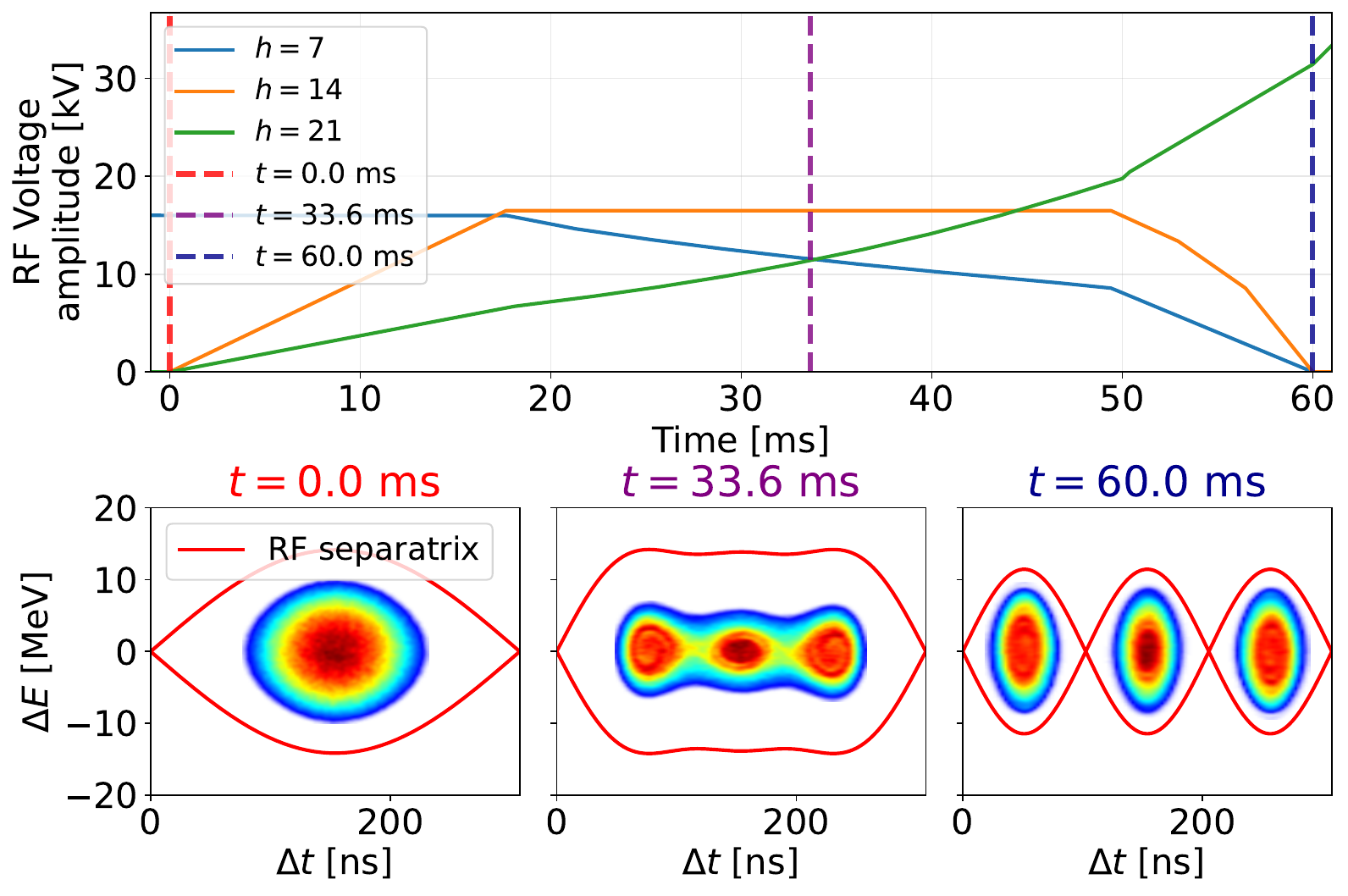}
    \caption{Overview of the triple splitting. The voltage programs of the three active rf systems at respective harmonics 7, 14, and 21 (current operational harmonics) are shown in the top plot. The longitudinal phase space of the particles are plotted for various times of the voltage program as indicated by the dashed lines (bottom). The phase space distributions are obtained with tracking with the simulation code BLonD~\cite{blond}.}
    \label{fig:triple}
\end{figure}

The manipulation is performed with the main rf system of the PS, consisting of several cavities tuneable from 2.8 to10~MHz to operate at different harmonics, specifically $h= 7,~14,~21$, at a stationary portion of the magnetic program, i.e. at constant beam energy. The manipulations are performed adiabatically, slowly with respect to the synchrotron frequency, in order to preserve the total emittance before/after the splittings~\cite{Garoby:rf_gymnastics}. A qualitative example of the voltages and the evolution of the longitudinal phase space as a function of time is shown in Fig.~\ref{fig:triple}.

The total rf voltage as a function of the longitudinal phase $\phi$ is given by,
\begin{equation}
    V_{\mathrm{rf}}(\phi) = \sum_{h} V_{h}\sin(h\phi + \phi_h),
\end{equation}
where the sum runs over the active harmonics $h$, and $V_h$ and $\phi_h$ are the peak voltage and phase offset of the corresponding rf system. Here $\phi$ is defined with respect to $h=1$, with the synchronous particle at $\phi=0$.
To obtain a symmetric splitting, the voltage at $h=14$ must be held in counter-phase with the $h=7$ and $h=21$, and the peak voltage must be large enough to drive particles out of the central bunch thereby populating the two outer bunches at $h=21$. To achieve this in operation, we tune the relative phasing of the rf voltages with the phase offsets $\phi_{14}$ and $\phi_{21}$, and scale the voltage amplitude of the $h=14$ program by multiplication with a dimensionless scaling factor, $\alpha_{14}$. All other parameters are held constant. Taking the $h=7$ phase as the reference ($\phi_7=0$), the total rf voltage for the triple splitting becomes
\begin{equation} \label{eq:v_rf_triple}
\begin{aligned}
    V_{\mathrm{rf}}(\phi;\, \phi_{14}, \phi_{21}, \alpha_{14}) = {}& V_{7}\sin(7\phi) \\
    &+ \alpha_{14}\,V_{14}\sin\!\left(14\phi + \pi + \phi_{14}\right) \\
    &+ V_{21}\sin(21\phi + \phi_{21}),
\end{aligned}
\end{equation}
where $V_{h}$ is the design amplitude of the $h$ voltage program, with $\alpha_{14}=1$ recovering the reference. The fixed phase offset of $\pi$ enforces the counter-phase condition of the voltage at $h=14$, and optimal splitting quality, in the absence of intensity effects, is obtained at $\phi_{14}=\phi_{21}=0$ and $\alpha_{14}=1$.

Deviations from this optimum can be understood term by term (Eq.~\ref{eq:v_rf_triple}). The phases $\phi_{14}$ and $\phi_{21}$ shift their respective harmonics in $\phi$, displacing the bunches. The voltage at $h=14$ shapes the earlier stage of the splitting and the one at $h=21$ the later one (cf.\ the rf program in Fig.~\ref{fig:triple}). A phase error introduces an asymmetry in the early or late bunch profile evolution accordingly. The factor $\alpha_{14}$ scales the amplitude at $h=14$, setting how strongly particles are pushed from the center toward the outer bunches, thereby depleting or overpopulating the central bunch.

All three parameters are not equally sensitive with respect to the quality of the splitting. Manual optimization of the parameters reach an accuracy in the order of half a degree in $\phi_{14}$, few degrees in $\phi_{21}$, and within $\pm1\%$ with respect to the reference voltage in $V_{14}$.

\subsection{Measuring splitting quality}

\begin{figure}[!]
    \centering
    \includegraphics[
        width=1.0\linewidth,
        trim=0 2.0cm 0 2.6cm,
        clip
    ]{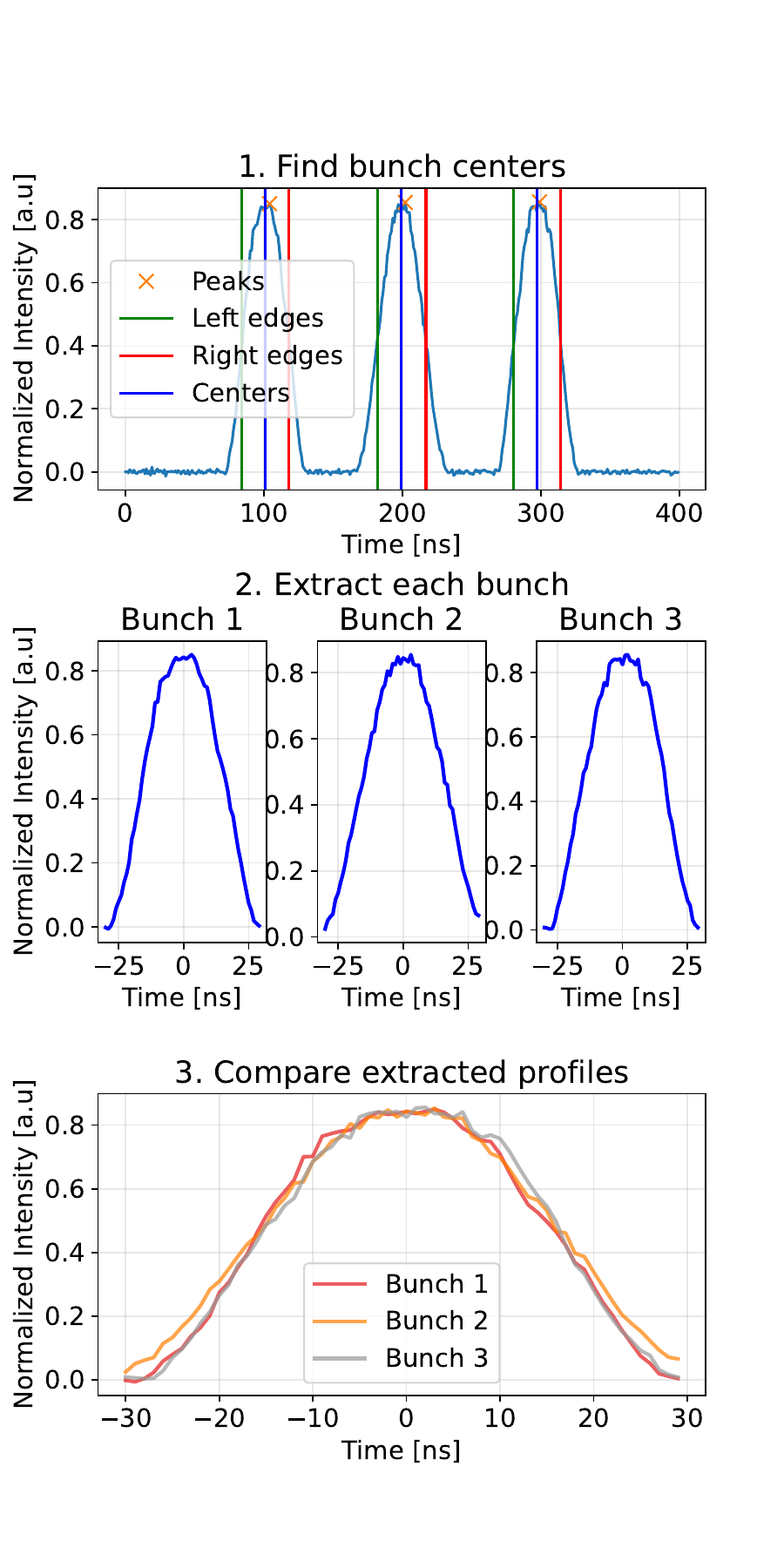}
   \caption{Illustration of the method of calculating the overall splitting quality loss. Firstly, the final bunch profile is analyzed, identifying individual bunches through peak detection. The center of each bunch is estimated to be in the middle of each half-max edge to the left and right. Each bunch is then isolated (middle plot), and overlayed (bottom plot). The MSE between all bunches is computed using Eq.~\eqref{eq:loss_tri} to assess the splitting quality.}
   \label{fig:init_triple_metric}
\end{figure}

To inform either an RL agent or optimizer of the current state, a measure of splitting quality is necessary. This can be achieved through analysis of the frequency distribution of the final bunch train using a discrete Fourier transform~\cite{Hancock}, or computing the difference in bunch lengths/intensities at the end of the splitting. In this work a new analysis is proposed which compares the entire bunch shape. It incorporates both bunch intensities and lengths into a single figure of merit through direct comparison of individual bunch profiles. The construction of the general quality metric used for the triple splitting is illustrated in Fig.~\ref{fig:init_triple_metric}. %

The bunch profile after the splitting process is analyzed, and each bunch center is extracted, which allows to separate the bunch profiles, and then compare them with each other through a Mean Squared Error (MSE),

\begin{equation}
\label{eq:mse}
    \mathrm{MSE}(y,\hat{y}) := \frac{1}{N}\sum_{i=1}^{n}(y_i - \hat{y_i})^2.
\end{equation}

The quality loss, $L_Q$, is the arithmetic average of the three MSEs:
\begin{equation}
\begin{aligned} \label{eq:loss_tri}
    L_Q(\phi_{14},\phi_{21},\alpha_{14}) = \\
     & \hspace{-1.9cm} \frac{\mathrm{MSE}(b_1,b_2) + \mathrm{MSE}(b_1,b_3) + \mathrm{MSE}(b_2,b_3)}{3}.
\end{aligned}
\end{equation}

\begin{figure}[!]
    \centering
    \includegraphics[width=1.0\linewidth]{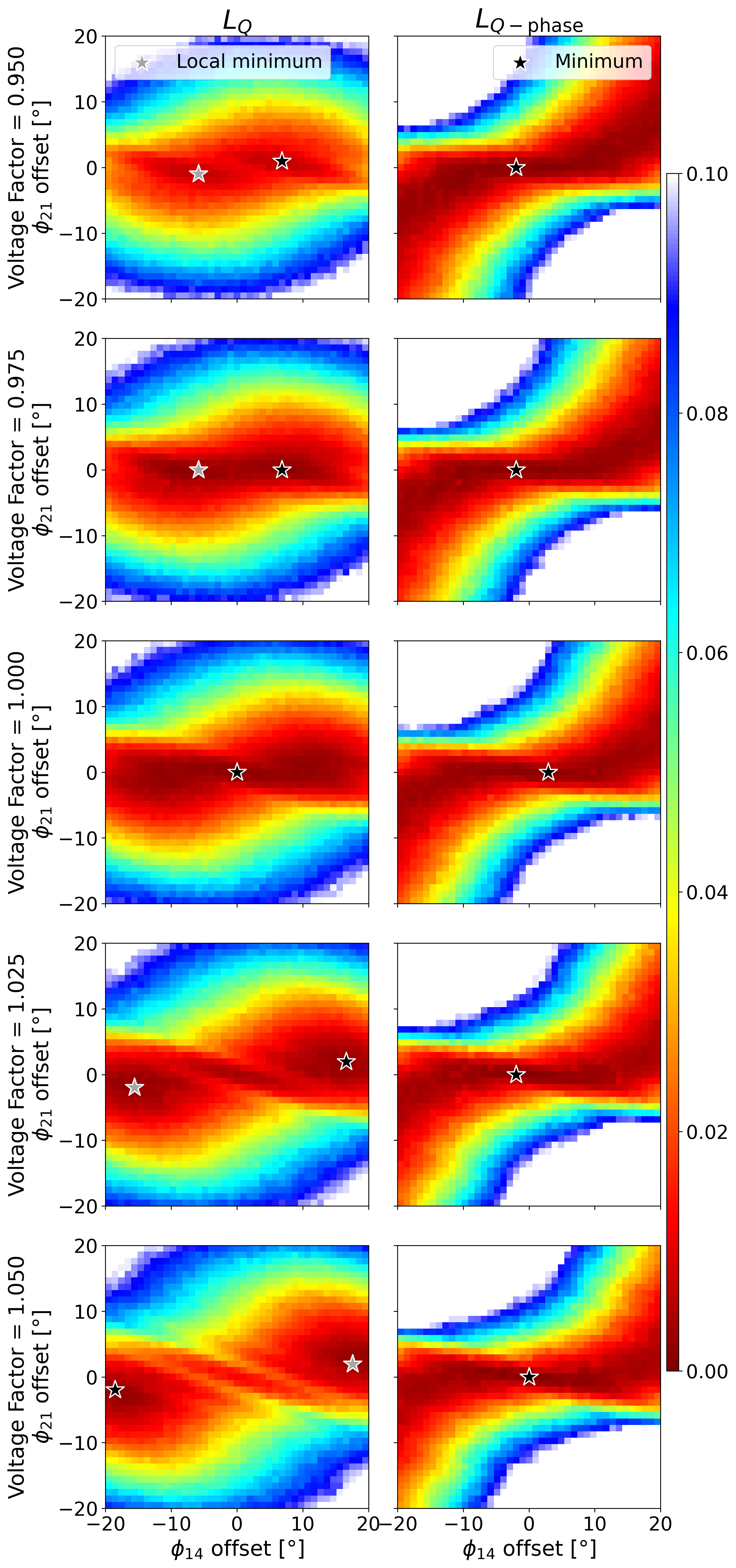}
    \caption{Scan of $L_Q$ (left column) and $L_{Q-\mathrm{phase}}$ (right column) for varying phase errors $\phi_{14}$ and $\phi_{21}$, based on BLonD simulation data for different normalized voltages (top to bottom). The voltage factor is multiplied to the reference voltage of the $h=14$ cavity. For loss quality $L_Q$, the local minima change significantly with varying voltage factors. The minima correspond to different asymmetric solutions of either equal final intensities or bunch lengths. The global minimum over all voltage factors is however still at zero phase errors and voltage factor equal to 1.0, as expected. The $L_{Q-\mathrm{phase}}$ scan on the right shows only minor dependency of the loss landscape with voltage, demonstrating that voltage and phase adjustments are approximately orthogonal in this case.
}
    \label{fig:loss_scan_combined}
\end{figure}

Here $b_1,$ $b_2,$ and $b_3$ are the individual isolated bunch profiles after the triple splitting. Minimizing $L_Q$ leads to minimal differences between bunch profile shapes. This implicitly forces the algorithm to optimize both for bunch intensity and length. However, as previously observed in~\cite{lasheen:ipac2021-wepab244}, optimizing all three parameters at once leads to a complex optimization landscape. This is further illustrated through scans of phase offsets for different fixed voltage factors in the left column of Fig.~\ref{fig:loss_scan_combined}. Nevertheless, this metric still provides a good representation of the overall splitting quality.

\subsection{Decoupling of phase and voltage parameters in triple splitting loss function}
\label{sec:decoupling_phas_volt}

Taking inspiration from operational experience with manually tuning the splitting, the parameters can be decoupled into sequential rf phase and voltage optimization steps. This has multiple benefits. It reduces the number of unnecessary changes to system settings for pure phase or voltage errors, and it enables faster iteration between methods due to the reduced action space of each problem.

For the optimization of rf phases, most variations will lead to a difference in the amount of particles in the outer two bunches regardless of the voltage setting. However, if the rf phases are already aligned, the two outer bunches will remain perfectly symmetrical regardless of the voltage amplitude of the $h=14$ cavity and only their relative size compared to the center bunch will change. By creating a loss function comparing only the outer two bunches with each, the phases can be initially tuned independently of voltage errors. The phase-loss quality is defined by $L_{Q-\mathrm{phase}}$:

\begin{equation}
\label{eq:loss_phase_tri}
    L_{Q-\mathrm{phase}}(\phi_{14},\phi_{21},\alpha_{14}) = \frac{\mathrm{MSE}(b_1,b_3)}{2},
\end{equation}

where $b_1$ and $b_3$ represent the outer two bunches after the triple splitting, and $\mathrm{MSE}$ as in Eq.~\eqref{eq:mse}. A scan of the phase-loss quality for different phase offsets and different voltage factors in Fig.~\ref{fig:loss_scan_combined} visualizes the new loss function. As desired, the loss is now independent from the voltage. This is valid for rf phase errors in the relevant optimization range.

\begin{figure}
    \centering
    \includegraphics[width=84mm]{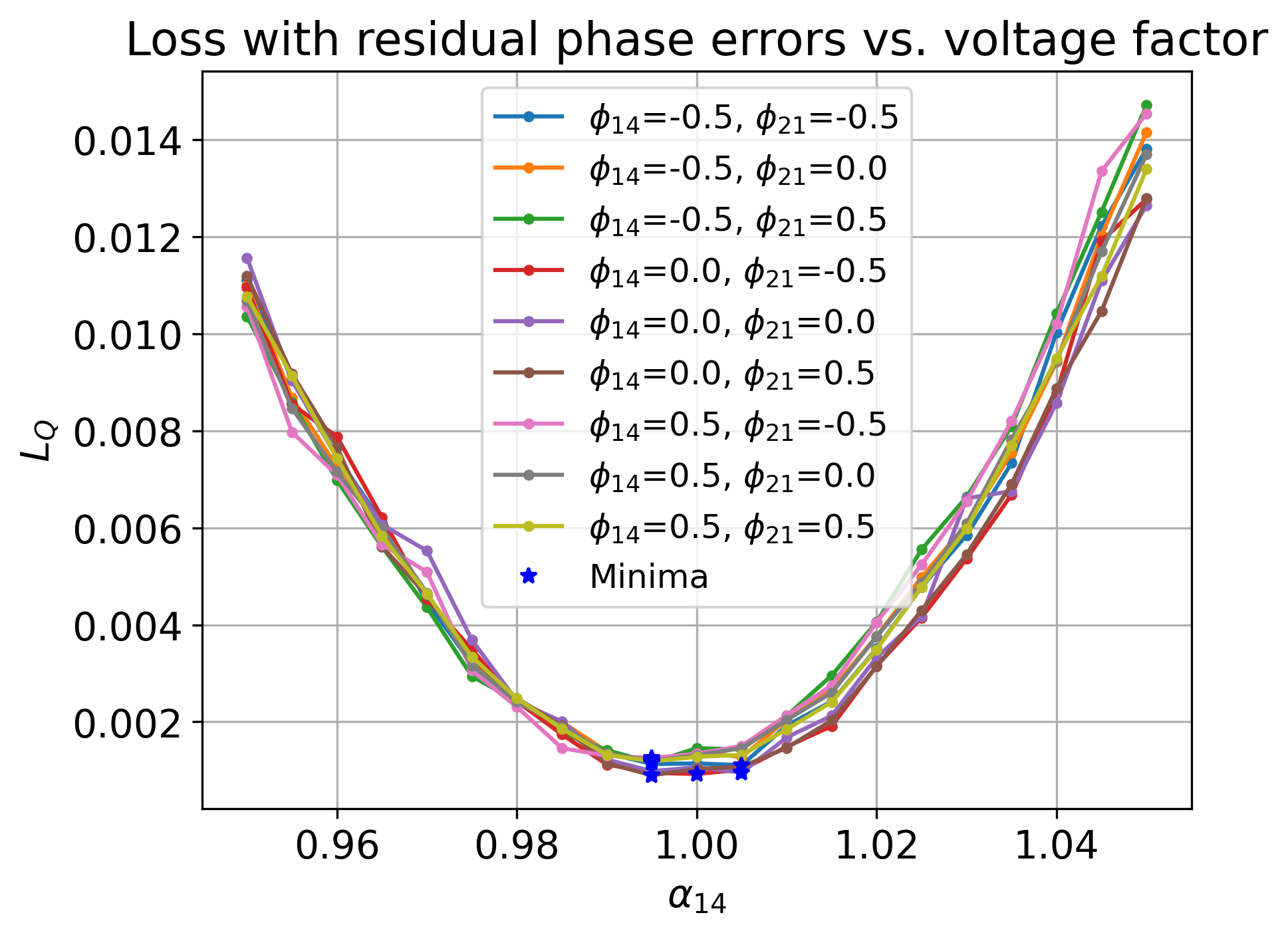}
    \caption{Simulated scan of phase loss quality $L_Q$while varying the voltage error (colored lines). If rf phases are optimized first, this loss function remains to be optimized for the voltage agent.}
    \label{fig:voltage_loss}
\end{figure}

With phases assumed to be corrected, the voltage optimization is reduced to a univariate problem. Re-using the loss function defined in Eq.~\eqref{eq:loss_tri} and scanning the intensity, while fixing the phases close to optimal, shows the loss is now approximately parabolic, even for small residual phase errors, as displayed in Fig.~\ref{fig:voltage_loss}.

\section{Simulation and Machine learning framework}

After preliminary testing of different RL algorithms, we converged on applying Soft Actor-Critic (SAC)~\cite{SAC}s. This was chosen due to its high-performance on common RL benchmarks and relatively small need of hyperparameter tuning. It is a model-free algorithm, and is among the most sample-efficient ones. It shares many similarities with other model-free RL algorithms, like Twin Delayed Deep Deterministic Policy Gradient (TD3)~\cite{td3}, which have efficient in solving accelerator control problems~\cite{sample-eff-RL}. It also showed the most promise of the methods used in a previous work on optimizing the longitudinal splittings of the PS~\cite{wulff_2021_0gmr6-bqj11}. The training was performed with simulated data, as the repetition rate of the PS would be too low (30~s to 1~minute for experimental beams) to train agents efficiently. The Stable Baselines 3~\cite{stable3} SAC implementation is used, and the environments were originally developed within the Open AI gym framework~\cite{DBLP:journals/corr/BrockmanCPSSTZ16} and later converted to Gymnasium~\cite{towers2024gymnasium}. A separately trained feature extractor was implemented using PyTorch~\cite{Ansel_PyTorch_2_Faster_2024} as a basic CNN. The feature extractor was investigated primarily as it's own tool, attempting to predict phase and voltage errors for real data, which would even enable direct feed-forward correction.

\subsection{Simulating the datasets}

For both the supervised learning and RL approach, a simulated dataset was produced for training. The tracking code BLonD was used to simulate the triple splitting~\cite{blond}. To reduce simulation time, collective effects were not considered. Although transient beam loading is a relevant intensity effect, in average it only perturbs the triple splitting like a relative phase error. Hence, the model should be able to generalize from data without the effect. Disregarding these effects also allows the dataset to be agnostic to the absolute intensity of the simulated bunches, enabling models trained on it to generalize over different beam production schemes.

Despite these simplifications, simulation of a single sample takes about 3 to 4 minutes. It was therefore decided to pre-simulate a large amount of labeled data, that could later be accessed quickly when required for training. This comes with the drawback of requiring interpolations to be made to enable continuous scanning of parameters at training time. 

A grid scan of the three parameters of the triple splitting was conducted, varying phase errors $\phi_{14},~\phi_{21}~\in~[-20, 20]$ in steps of 1 degree, and voltage factor $\alpha_{14}~\in~[0.95,1.05]$ ($\pm5\%$ compared to a reference voltage) in steps of 0.005. An additional fine scan of phase errors was conducted in the same range, varying phase by 0.025 degrees while keeping the voltage factor constant at $\alpha_{14}=1.0$. In total, about 60k unique setting combinations were simulated.

\subsection{Domain randomization/adaption}
\label{sec:domain-rando}

The result of the BLonD simulations, while very accurate, do not present some features only obtained in measurements. Indeed, no modeling is done of the measurement systems transfer functions, no measurement noise is present in the final bunch profiles, and cycle-to-cycle beam parameter variations from upstream accelerators are not considered. This makes the simumlation to real transfer (often abbreviated as sim2real\footnote{Sim2real refers to the challenge of transferring an agent's performance from training in a simulated environment to operation in the real environment.} in RL contexts) of any model trained on pure simulation data difficult. To obtain a more generalized model, ready for transfer to the PS, we apply domain randomization techniques specifically engineered to adapt the simulation to resemble a real measurement, mimicking shot-to-shot variations through artificial randomness. This is done by three main operations:

\begin{enumerate}
    \item Normalization of the bunch profiles, to remove dependency on real number of particles in each bunch.
    \item Adding estimated measurement noise, modeled as purely Gaussian. To characterize typical noise levels, measurements of the triple splitting were made and areas with no particles were sampled to estimate the mean ($\mu$) and standard deviation ($\sigma$).
    \item Adding longitudinal shifts (modelling small timing shifts). The bunch centers were artificially offset by shifting the entire matrix of bunch evolutions along the time axis, with the size of the shift being sampled from a uniform distribution in the range $[-6, 6]$~nanoseconds. This emulates a beam phase jitter with respect to the trigger.
\end{enumerate}

\begin{figure}[!h]
    \centering
    \includegraphics[width=1.0\linewidth]{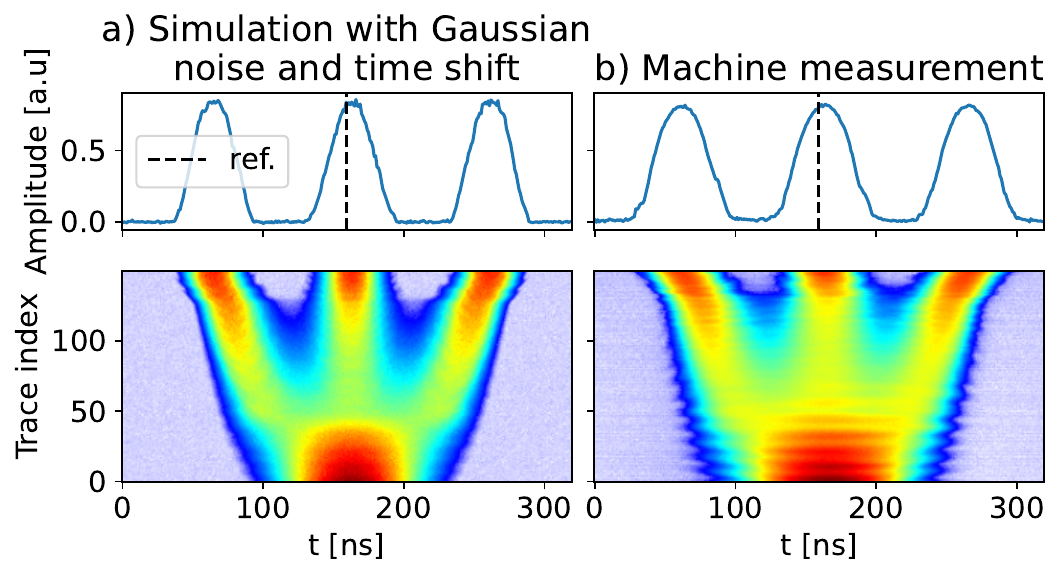}
    \caption{Simulated and measured triple splitting evolution in the PS. In the bottom heat-map plot, the longitudinal profile of the beam is plotted in each row every 185 turns in the PS, resulting in the evolution over time from the bottom row to the top row of the image. This type of visualization is often referred to as a mountain range plot. The final profile is additionally displayed in the top part of the figure, showing the final state of the beam after the manipulation is complete. The dashed reference line shows the center bin of the acquisition. Note that in this particular measurement, there is a misalignment of the center, and oscillations from earlier stages of the beam are present.}
    \label{fig:sim-domain-rando}
\end{figure}

An example of the simulated full bunch profile evolution and final profile before and after application of the domain randomization is shown in Fig.~\ref{fig:sim-domain-rando}.

It can be noted that the oscillations visible in the real measurement of Fig.~\ref{fig:sim-domain-rando} b) are not a feature of the triple splitting, but stem from earlier parts of the beam generation. The variation in initial conditions are not modeled in our current dataset or domain randomization, but could be a future extension to provide an even more general model.

\subsection{The feature extractor}

\begin{figure}[!h]
    \centering
    \includegraphics[width=81mm]{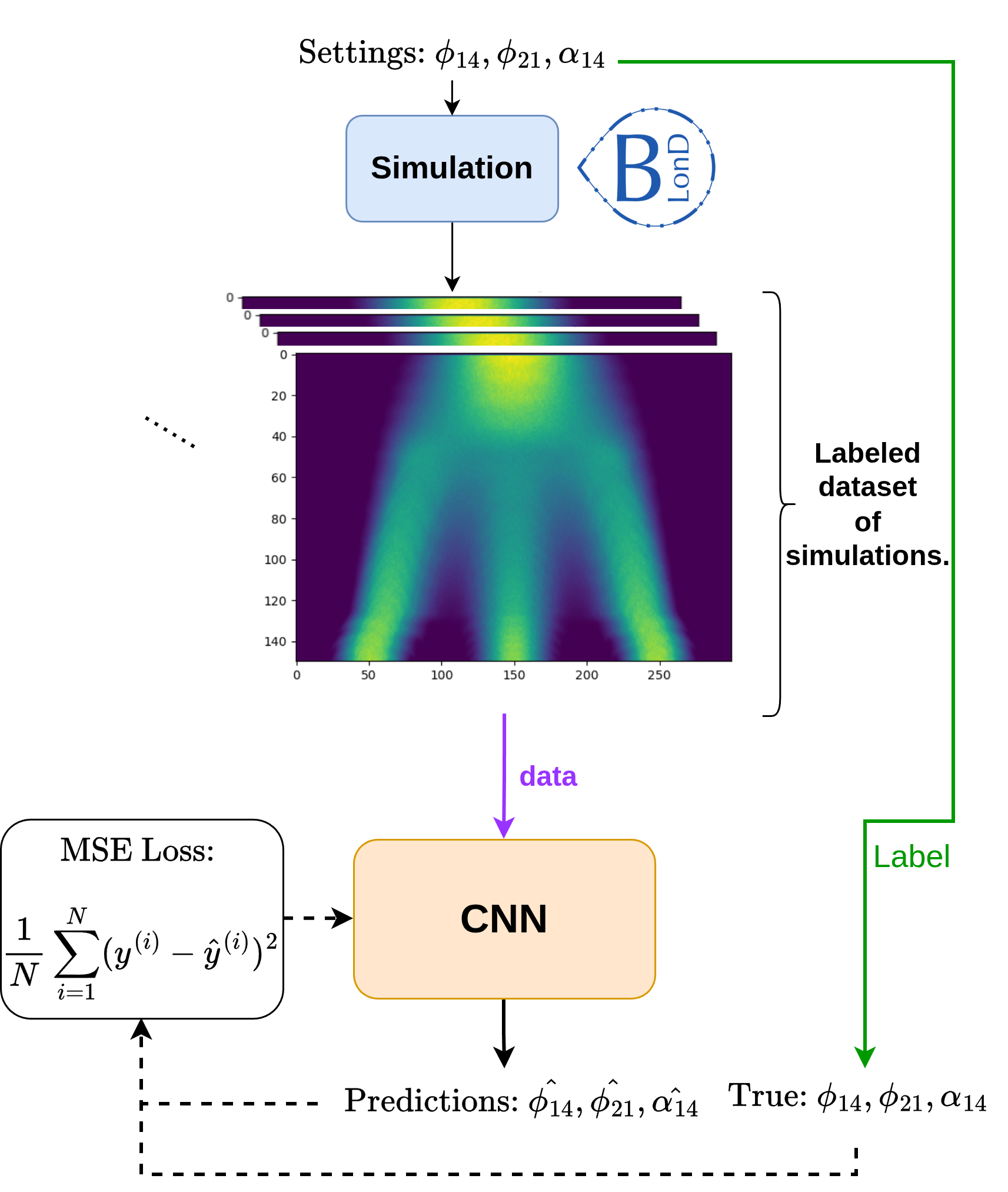}
    \caption{Supervised training loop for the feature extractor CNN. A scan of varying phase and voltage (factor) settings is performed to build a labeled dataset of bunch evolutions using BLonD. The simulated bunch evolution matrices are used as input to the CNN, which is tasked with predicting the settings initially used for their simulation.}
    \label{fig:feat_extractor_training}
\end{figure}

\begin{figure*}
    \centering
    \includegraphics[width=1.0\linewidth, clip]{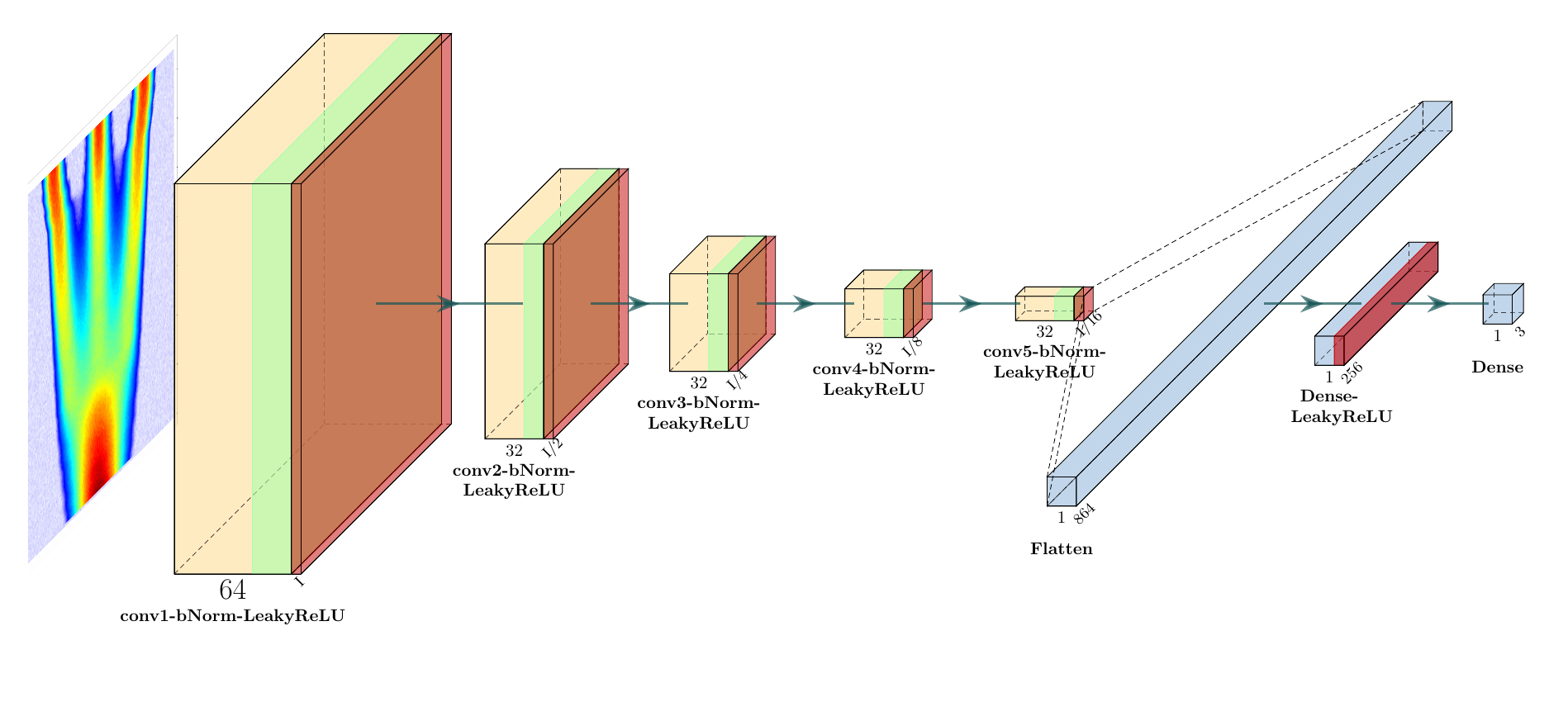}
    \caption{Architecture of the feature extractor neural network. Simple CNN of five consecutive 2D-convolutional blocks, where each block is composed of a strided convolutional layer ($\mathrm{stride}=2$), a batch normalization layer, and a leakyReLU activation layer (leading to the shorthand names convX-bNorm-LeakyReLU used in the figure). The convolutional blocks are followed by two fully connected layers compressing the features into three output parameters: absolute phase errors in $\phi_{14}$ and $\phi_{21}$, and voltage factor $\alpha_{14}$.}
    \label{fig:feature_extractor_arch}
\end{figure*}

Once the complete simulated dataset is generated, a first model can already be designed to extract the phase and voltage offsets applied as input to the simulation. This effectively reverses the problem which cannot be easily done analytically. Such a model is named the feature extractor, as it learns a feature map correlating bunch evolution matrices with specific system settings (features). A schematic of the training procedure is illustrated in Fig.~\ref{fig:feat_extractor_training}, and the architecture of the feature extractor is shown in Fig.~\ref{fig:feature_extractor_arch}. 

The model acts as an image encoder, following a traditional CNN architecture. Rather than design a bespoke network, we assembled well-established components and kept the configuration that performed well on our data. It consists of five layers of strided convolutions ($\mathrm{stride}=2$)\footnote{$\mathrm{Stride}>1$ means each layer downsamples the spatial dimension of the input. The filter for the donwsampling is learned through the layers weights and has been shown to not lose accuracy compared to traditional pooling layer's~\cite{Springenberg2014StrivingFS}.}, each followed by batch normalization~\cite{pmlr-v37-ioffe15} layers and leaky Rectified Linear Unit (ReLU) activation functions~\cite{Maas2013RectifierNI}. Two considerations fix the overall shape of the network. The depth of five layers allows the deeper ones to combine information from across the entire bunch-profile matrix, and the progressive downsampling from the strided convolutions systematically reduces the spatial dimensionality of the features. This restricts the parameter count of the fully connected layers, keeping the model compact. The final layers map the compressed features onto three continuous parameters representing the absolute phase errors in $\phi_{14}$, $\phi_{21}$ and voltage factor $\alpha_{14}$.

The network input is the full matrix of bunch profiles (Fig. \ref{fig:sim-domain-rando}, bottom), after passing through the domain randomization steps described in Section~\ref{sec:domain-rando}. The outputs are trained with a supervised MSE loss to reconstruct the settings used to generate each sample: i.e., the true absolute phase errors and voltage factor. The training used a 9/1 training/validation split of the data. 

The final model was trained for 43 epochs\footnote{In machine learning, an epoch refers to one full use of the entire dataset to update the model parameters.} and was stopped after reaching satisfactory performance on the simulation test set, with about 0.5~degrees of phase errors and about 0.3~\% in amplitude of the $h$=14 voltage factor. The achieved accuracy of the model with simulated bunch profiles is shown in Table \ref{tab:feat_extr_sim}, reaching an average absolute error smaller than 0.1 degrees for both phases and less than 0.2~\% of the reference voltage. This is better than the resolution of changes during manual optimization.

\begin{table}[]
    \centering
    \begin{tabular}{|c|c|c|}
    \hline
    \multicolumn{3}{|c|}{CNN Absolute prediction error (sim.)} \\
    \hline
    $\phi_{14}$ [deg] & $\phi_{21}$ [deg] & $1-\alpha_{14} [\%]$ \\
    \hline
    \hline
     $0.037\pm 0.031$ & $0.063\pm 0.051$ & $0.17\pm 0.13$ \\
    \hline
    \end{tabular}
    \caption{Performance of the trained feature extractor on the simulation test set. }
    \label{tab:feat_extr_sim}
\end{table}

Despite the promising results with simulated bunch profile data, the method did not transfer its prediction accuracy to measurements. Absolute prediction errors grew too large for one-shot corrections. The predicted phases were still within degrees of the true phase offsets, which meant they could be leveraged to improve initial conditions, but not fully reach target splitting quality (further explored in Sec. \ref{sec:models_in_operation}). This motivated searching for alternative methods for the finetuning of the splitting parameters.

\subsection{Reinforcement Learning environments}

\begin{figure}
    \centering
    \includegraphics[width=1.0\linewidth]{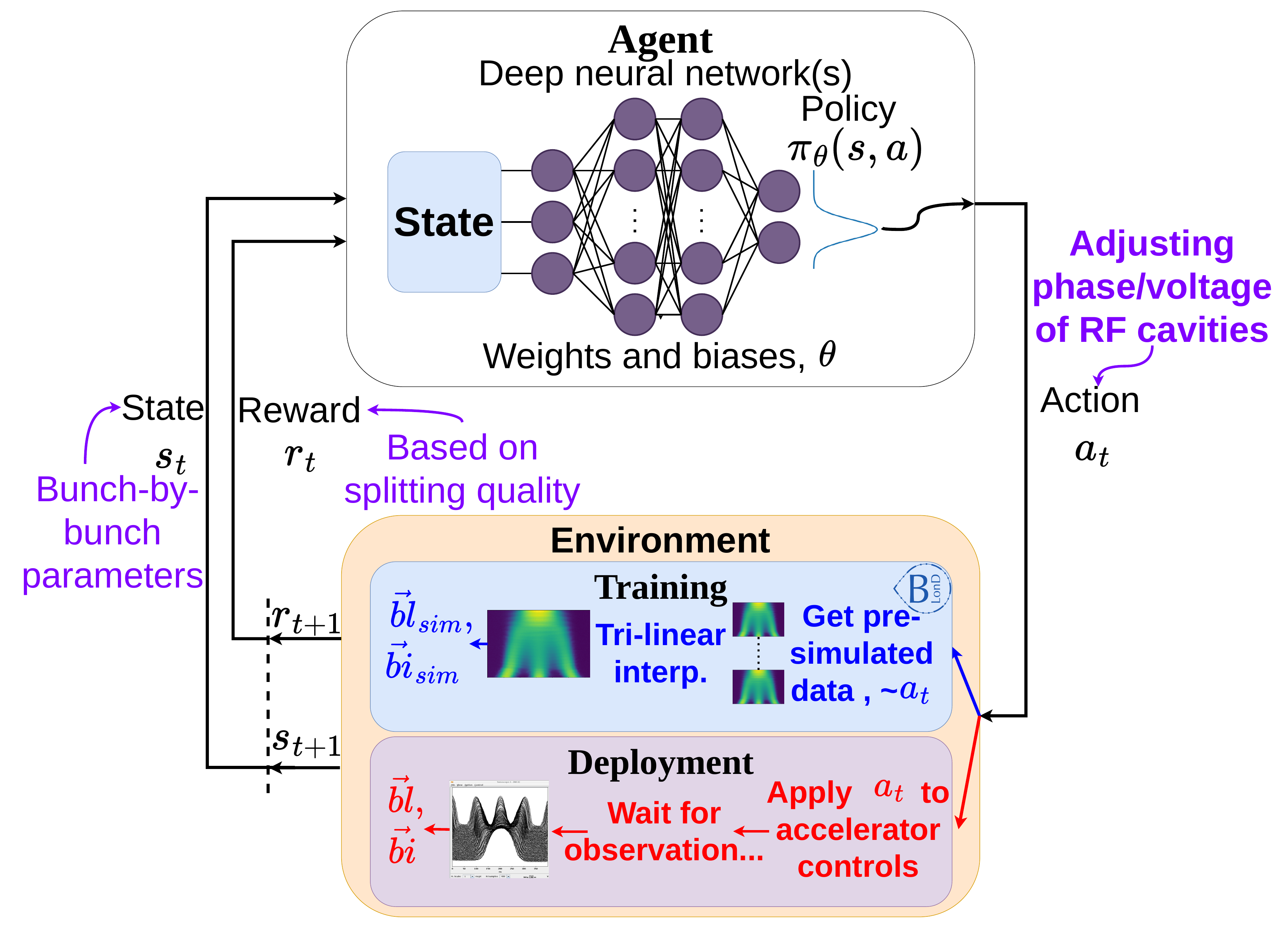}
    \caption{
    The agent-environment interaction loop for training and deployment. During training, the agent interacts with datapoints from a pre-simulated dataset. These datapoints are used to calculate a state ${s_t}$, which the agent uses to take an action, $a_{t}$. It then receives an updated state ($s_{t+1}$) and a reward $r_{t+1}$ from the environment, which is maximized by the RL algorithm. During inference (deployment), the exact same interface is used as in training, with the only difference being that $s_{t}$ is calculated based on measurements of beam profiles rather than pre-simulated samples.}
    \label{fig:agent_environment_loop}
\end{figure}


As a complementary approach, we investigated applying RL agents for the optimization, with the motivation that a learned action behavior (or policy) may transfer more smoothly to the accelerator. Through RL training, an agent learns which actions to take relative to the current state of the splitting. This means it can learn strategies of actions in a simulated environment, that could transfer to the real one (even if absolute differences remain). Two separate environments were required for the implementation of the agents: a simulation-based one for training, and one connected to the accelerator for deployment. Both domains are defined with the same action and observation spaces, but differ in how they acquire their observations. In training, we profit from the pre-simulated data for training as for the feature extractor. For an arbitrary setting within the range of the dataset, a tri-linear interpolation of the nearest data points is performed. The same domain randomization techniques as for the feature extractor are applied to enhance the robustness of the final agents. The pre-simulated data plus interpolation acts as an efficient surrogate model, enabling fast training. An episode of training terminates either when the optimization is complete or 100 steps have been reached. In the real environment of the PS, the actions requested by the agents are sent directly through the accelerator controls system to the hardware, and the observation is derived from the measurement during the next acceleration cycle. The agent-environment interaction loop is illustrated in Fig. \ref{fig:agent_environment_loop}.

\subsubsection{Action space}

By segmenting the optimization of the manipulation into a phase and voltage step, two separate training environments with varying action spaces are created. Actions are normalized to the interval $[-1, 1]$, preparing the action spaces $A_{\phi}$ and $A_{\alpha_{14}}$:

\begin{equation}
    A_{\phi} = \{a = [a_{\phi14},\ a_{\phi21}] \mid a_{\phi14},\ a_{\phi21} \in [-1, 1]\},\\\\
\end{equation}
\begin{equation}
    A_{\alpha_{14}} = \{a = [a_1] \mid a_1 \in [-1, 1]\}.\\\\
\end{equation}

The choice of normalization factor for both phases and voltage actions sets the maximum step size of the agent during optimization. They are hyperparameters of the environment that needs to be set before training. The step sizes used in the operational models are presented in Sec.~\ref{sec:RL_training_details}.

\subsubsection{Observation}

The observations given to the RL agents are derived from the final bunch profile only, in contrast to using all traces of the bunch evolution as for the feature extractor. This choice reduces the complexity and dimensionality of each observed input. By avoiding high-dimensional raw profiles, we mitigate the curse of dimensionality~\cite{bellman1966dynamic}, which is particularly critical for the sample efficiency of model-free algorithms like SAC. This was found to aid in the convergence stability and speed of the RL training. Additionally, restricting the observation to these key derived parameters acts as a physics informed filter. It effectively removes low-level instrumentation noise and simulator-specific artifacts, and prevents the RL agent from overfitting to the simulation data, thereby improving generalization and robustness when transferring the trained policy to the real accelerator.

From each datapoint, the final profile is extracted after the manipulation is completed. The bunch-by-bunch lengths and intensities are normalized by dividing by the max bunch length or intensity respectively, and then centered around zero by subtracting their mean. These relative representations of the measurements are given as inputs to the RL agent.

For the two respective agents of the triple splitting, the observations are defined as follows:

\begin{equation}\label{eq:phase_obs}
    O_{\phi} = \{o = [\tau_1, \tau_3, I_1, I_3] \mid \tau_1, \tau_3, I_1, I_3\in [-1,1]\},
\end{equation}
\begin{equation}
\begin{aligned} \label{eq:volt_obs}
    O_{\alpha_{14}} = \\
    & \hspace{-1.5cm} \{o = [\tau_1, \tau_2, \tau_3, I_1, I_2, I_3] \mid \tau_1, \tau_2, \tau_3, I_1, I_2, I_3\in [-1,1]\}.
\end{aligned}
\end{equation}

Here, $\tau_n$ and $I_n$ represent the relative bunch length and intensity respectively of bunch number $n$.

\subsubsection{RL reward}

\begin{figure}
    \centering
    \includegraphics[width=\linewidth]{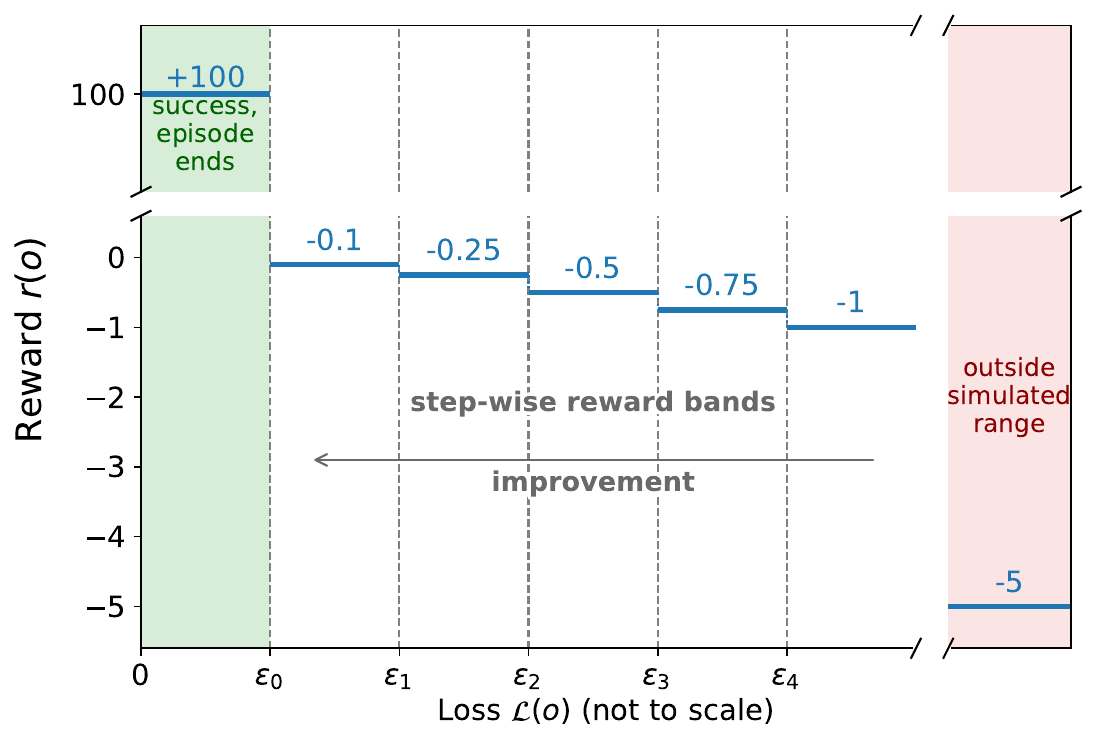}
    \caption{Schematic illustration of the reward function shape of the RL agents. The horizontal axis shows the loss $\mathcal{L}(o)$ (not to scale), with the thresholds $\epsilon_{0\text{--}4}$ marked by dashed lines. When the loss falls below $\epsilon_0$ (green shaded region), the target is reached: the agent receives a reward of $+100$ and the episode terminates. Above $\epsilon_0$, the constant negative reward per step decreases in bands towards the target, providing a direction of improvement. The out-of-range penalty of $-5$ (red shaded region, beyond the axis break) is not triggered by a loss threshold, but whenever any phase or voltage setting leaves the simulated parameter range.}
    \label{fig:reward}
\end{figure}

The reward functions are based on the loss functions introduced in Eq.~\eqref{eq:loss_tri} and Eq.~\eqref{eq:loss_phase_tri} respectively. Several variants were tested initially, and a simple sparse, stepwise reward converged to the optimum faster than comparable continuous rewards. Since this simplified approach already gave satisfactory performance (see Table~\ref{tab:sim-rl-agents}), it was adopted for the final agents. Each reward is a step function of its loss, as illustrated in Fig.~\ref{fig:reward}.

The thresholds $\epsilon_{0\text{--}4}$ segments the loss into bands, with the constant negative reward per step shrinking from band to band towards the target. The lowest threshold, $\epsilon_0$, was set to match our target parameters. Crossing it yields a large positive reward and terminates the episode. The remaining thresholds, together with their number and the associated reward values, were tuned empirically. The negative per-step rewards encourage fast convergence, while their band-wise decrease towards the target provides a direction of improvement. Finally, if the agent strays outside the region covered by the dataset, a larger penalty is assigned and a dummy observation is returned indicating which parameters are out of range, which keeps training within the relevant parameter region. 



\subsection{RL training details}
\label{sec:RL_training_details}

The two RL agents are named SAC-phase and SAC-volt. Both agents implement the same architecture in their policy network with the exception of the input and output layers, which change with the observation and action spaces for the different tasks. We follow the original SAC architecture of two fully connected layers of 264 nodes connecting input/output~\cite{SAC}. The entropy coefficient was fixed during training to $0.2$, as it was found to provide faster convergence in our environment then leaving it as a learned parameter.

The SAC-phase agent action space is normalized with maximum step size of 10 degrees in both phases, while the SAC-volt agents actions are normalized to a maximum of 2\% of the reference voltage per step.

Both agents are trained separately. For the SAC-phase agent, the initial state for each episode is generated with a random phase and voltage error sampled from two uniform distributions, corresponding to phase errors in the range of $\pm10$~degrees and voltage errors of $\pm5\%$ compared to the reference voltage.

For the SAC-volt agent the phase is assumed to be approximately optimal in the initial state. Phase errors are only the residual errors left after the phase optimization is complete. To model this, the voltage environment samples its initial phase error from a uniform distribution in the range $\pm1$~degree, allowing the voltage agent to learn how to act even in scenarios with a minor residual phase error remaining after the phase optimization, increasing robustness in real scenarios. The voltage error is initialized in the same manner as for SAC-phase.

\begin{figure}[t!]
    \centering
    
    \subfloat[
        RL agent optimizing phases in the simulated environment.
        \label{fig:performance-sim-phase}
    ]{
        \includegraphics[width=\linewidth]{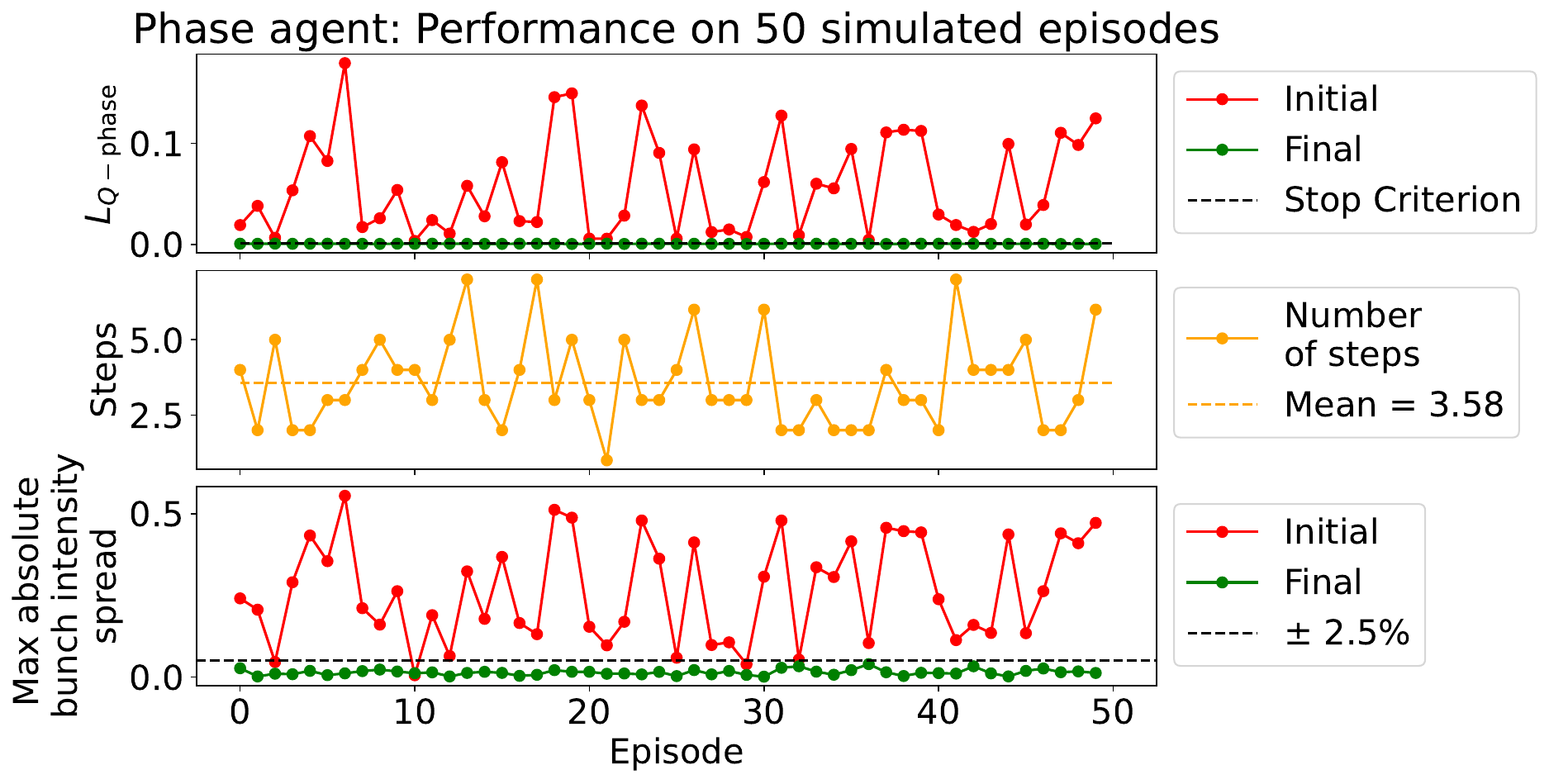}
    }
    
    \vspace{0.5cm} 
    
    \subfloat[
        RL agent optimizing voltage in the simulated environment.
        \label{fig:performance-sim-volt}
    ]{
        \includegraphics[width=\linewidth]{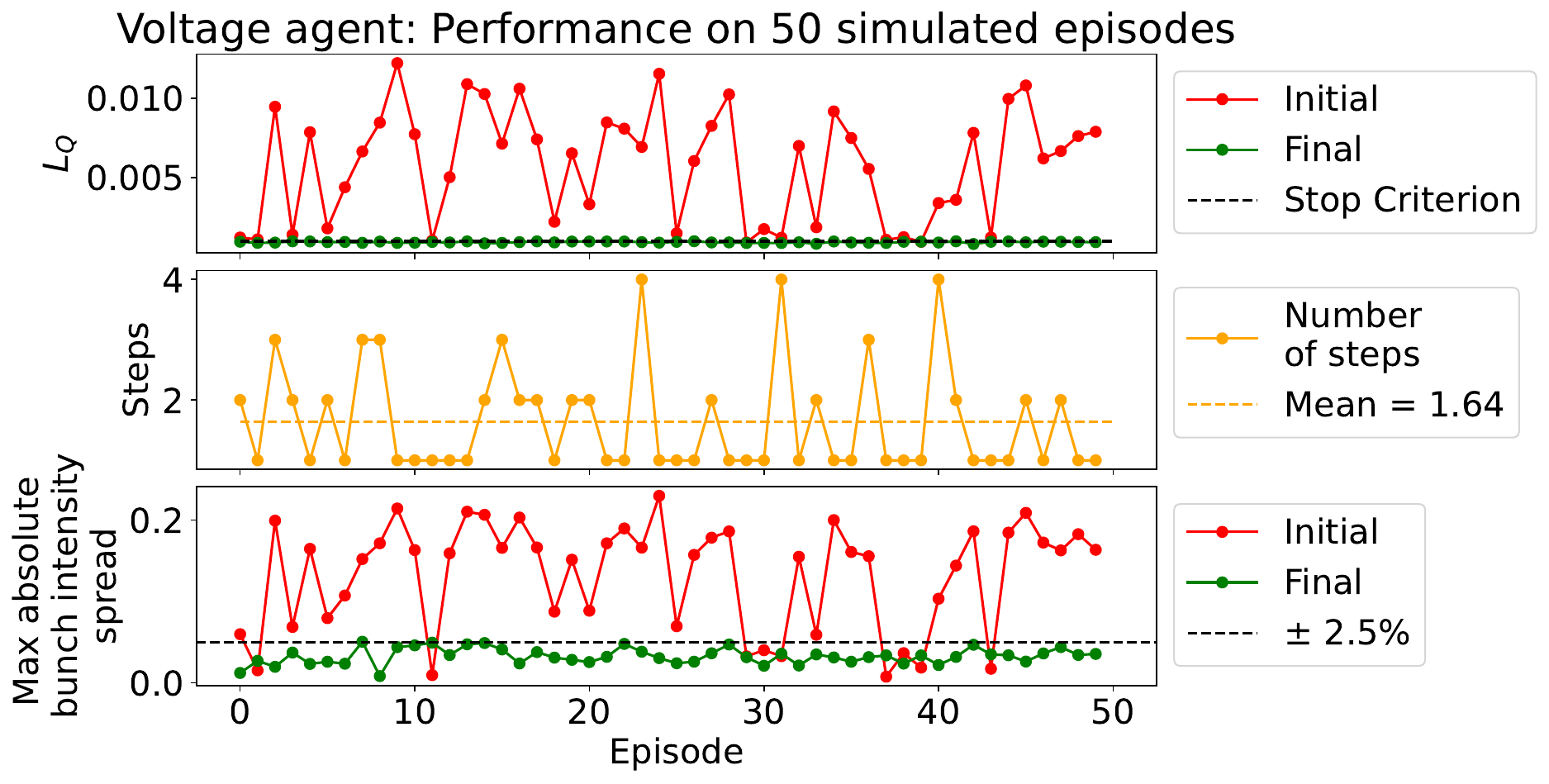}
    }
    
    \caption{
    Performance of trained RL agents in the simulated environment for phase (a) and voltage (b) optimization tasks. Quality improvement is illustrated in top and bottom plots, and number of steps to target in the middle. The additional metric of maximum absolute bunch intensity difference is shown in the bottom plot. In (a), the maximum absolute bunch intensity difference is computed using only the outer two bunches, while in (b) all final bunches are considered. The maximum absolute bunch intensity difference remains below $\pm2.5\%$ relative to the mean intensity. These results remain far below the upper acceptance limit for the final bunch train to the LHC of $\pm10\%$, which we remain far below.
    }
    \label{fig:performance-sim-combined}
\end{figure}

Both agents were set to train for $10^5$ steps, but saturated in performance before reaching the total number of steps. The final models were chosen as the ones performing best during evaluation, and were trained for approximately $6\cdot10^4$ steps for SAC-phase, and $2\cdot10^4$ steps for SAC-volt. The performance evaluated in the simulated environment for 50 test episodes is summarized in Table \ref{tab:sim-rl-agents} and Fig.~\ref{fig:performance-sim-combined}. The performance was sufficient to warrant testing in the PS, as it was estimated to be at least twice faster than manual adjustment on average.

\begin{table}[]
    \centering
    \begin{tabular}{|c|c|c|c|c|}
    \hline
    Model & \makecell{Mean number\\ of steps} & \makecell{Minimum \\number of\\ steps} & \makecell{Maximum\\ number of\\ steps} & \makecell{Success\\ rate} \\
    \hline
    \hline
    Phase agent & $3.6\pm 1.34$ & 1 & 16 & 100\% \\
    \hline
    Voltage agent  &  $1.64\pm0.89$ & 1 & 4 & 100\% \\
    \hline
    \end{tabular}
    \caption{Performance of trained RL agents in simulation environments, evaluated on 50 episodes.}
    \label{tab:sim-rl-agents}
\end{table}

\section{Exploiting the models in operation}
\label{sec:models_in_operation}

The initial performance of the feature-extraction model and the RL agents was benchmarked and presented in~\cite{internal-note-jwulff}. The testing demonstrated that neither the feature-extraction model nor the RL agents were able to reliably optimize the triple splitting when applied independently.

The feature extractor, while effective in simulations, showed reduced accuracy with real beam measurements: voltage predictions were unreliable and phase estimates, while still within a few degrees of the true phase error, were not accurate enough to fully optimize the splitting. The RL agents could iteratively refine the phase and voltage settings to an optimized state, but only when launched near a suitable parameter set. Their reliance on final longitudinal profiles made them struggle in cases with large initial phase errors, only visible earlier during the bunch profile evolution.

\begin{figure}[!h]
     \centering
     \includegraphics[width=0.9\linewidth]{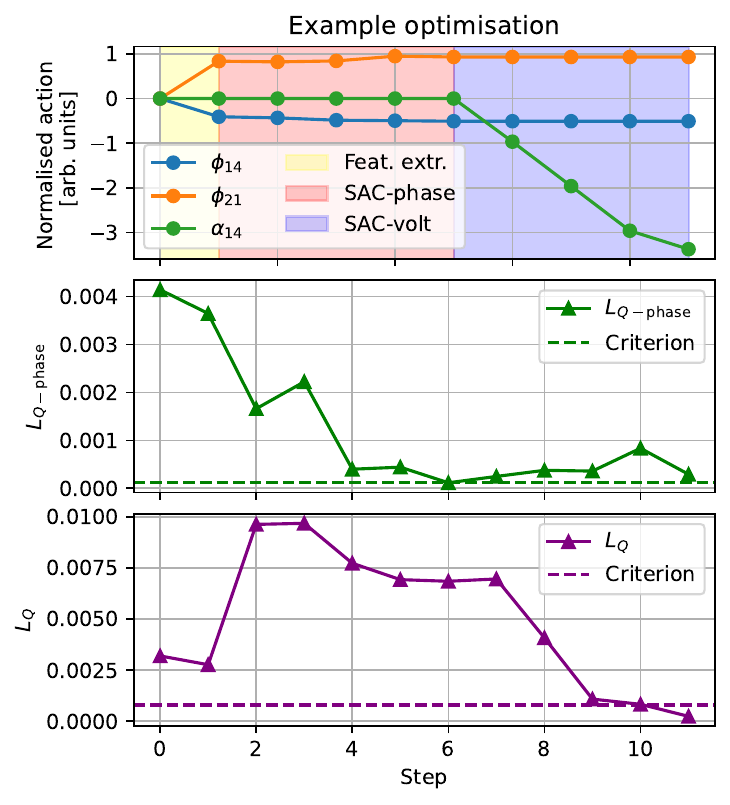}
     \caption{ Example episode of optimization with the on-demand script. The initial phases $\phi_{14}$ and $\phi_{21}$ were given an error of $+5$ and $-10$ degrees respectively, and the initial voltage was increased by $+7.5\%$ compared to the reference voltage.  In the top plot the normalized actions are shown, along with its corresponding model. In the middle and bottom plots, we see the evolution of the optimization losses for phase ($L_{Q-\mathrm{phase}}$) and for voltage ($L_Q$, overall quality) over the course of the episode~\cite{internal-note-jwulff}.}
     \label{fig:example_episode}
 \end{figure}

\begin{figure}[!h]
     \centering
     \includegraphics[width=0.9\linewidth]{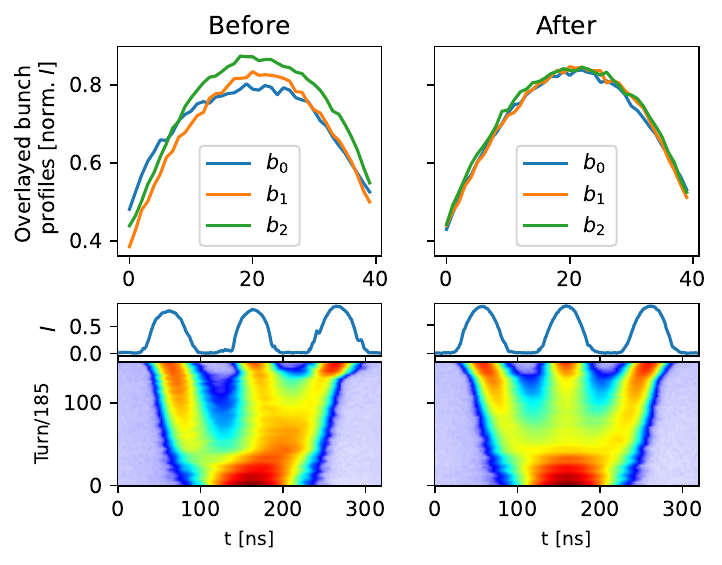}
     \caption{Initial and final bunch splitting characteristics for the example episode in Fig. \ref{fig:example_episode}. Before optimization, the bunch shapes after the splitting are mismatched and the overall bunch evolution is skewed. After optimization, the final bunch shapes align well and overall bunch evolution is symmetric~\cite{internal-note-jwulff}.}
     \label{fig:example_episode_before_after}
 \end{figure}

Both approaches exhibit complementary failure modes, motivating combining them into a sequential optimization scheme. The feature extractor was first applied to exploit the full bunch profile evolution information, providing an initial coarse phase correction leading to a more symmetric splitting. The phase- and voltage-optimization agents were then applied sequentially to refine the settings to operational accuracy. This combined method was deployed operationally in the PS as an on-demand script from 2023 to 2025 and consistently delivered fast, reliable optimization. An example of a full optimization episode is shown in Figure \ref{fig:example_episode}. The large phase correction predicted by the feature extractor in the first step is worth noting, followed by iterative refinement of the phases and voltage by the RL agents. A before and after comparison of the bunch profiles is shown in Fig. \ref{fig:example_episode_before_after}.

Several limitations in the beam observation systems of the PS prevented a fully autonomous implementation. In the second half of 2024, new hardware and accompanying acquisition software became available. This new system is referred to as the Longitudinal Beam Observation system (LBO)~\cite{LBO_amaury}. It features enhanced flexibility in acquisition, and allows for permanent signals to be setup on a beam-by-beam basis, enabling an extension of the on-demand scheme into a fully automatic and constantly monitoring implementation.

\subsection{A fully autonomous implementation}
\label{sec:fully_autonomous}

\begin{figure}
    \centering
    \includegraphics[width=1.0\linewidth]{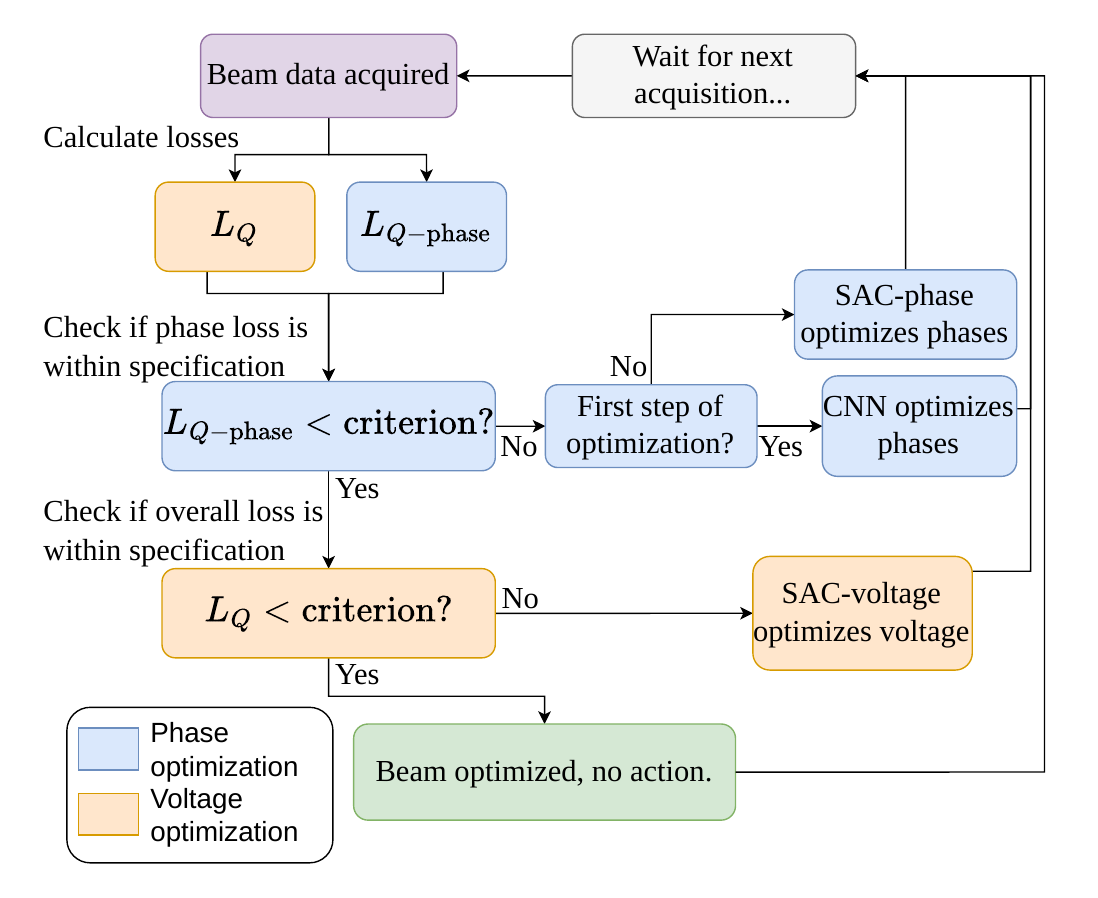}
    \caption{Decision tree for the autonomous version of the optimizer. After initialization, the system waits for the next beam observation to be acquired, calculates metrics, checks them against the set criteria in the control system, and acts only if necessary. Unlike the on-demand setup, the loop alternates freely between phase and voltage optimization.}
    \label{fig:continuous_monitoring}
\end{figure}

Once permanent beam observations became available, the existing on-demand optimizer was adapted to a continuously running version without major architectural changes. The optimization loop was modified to trigger based on automatic data refresh with the cycle of the accelerator. Actions are taken whenever the accompanying loss degrades above a set threshold, prioritizing optimization of phase before voltage. The autonomous controller loop implementation is illustrated in Fig.~\ref{fig:continuous_monitoring}.

\begin{figure*}[!ht]
    \centering
    \includegraphics[width=1.0\linewidth]{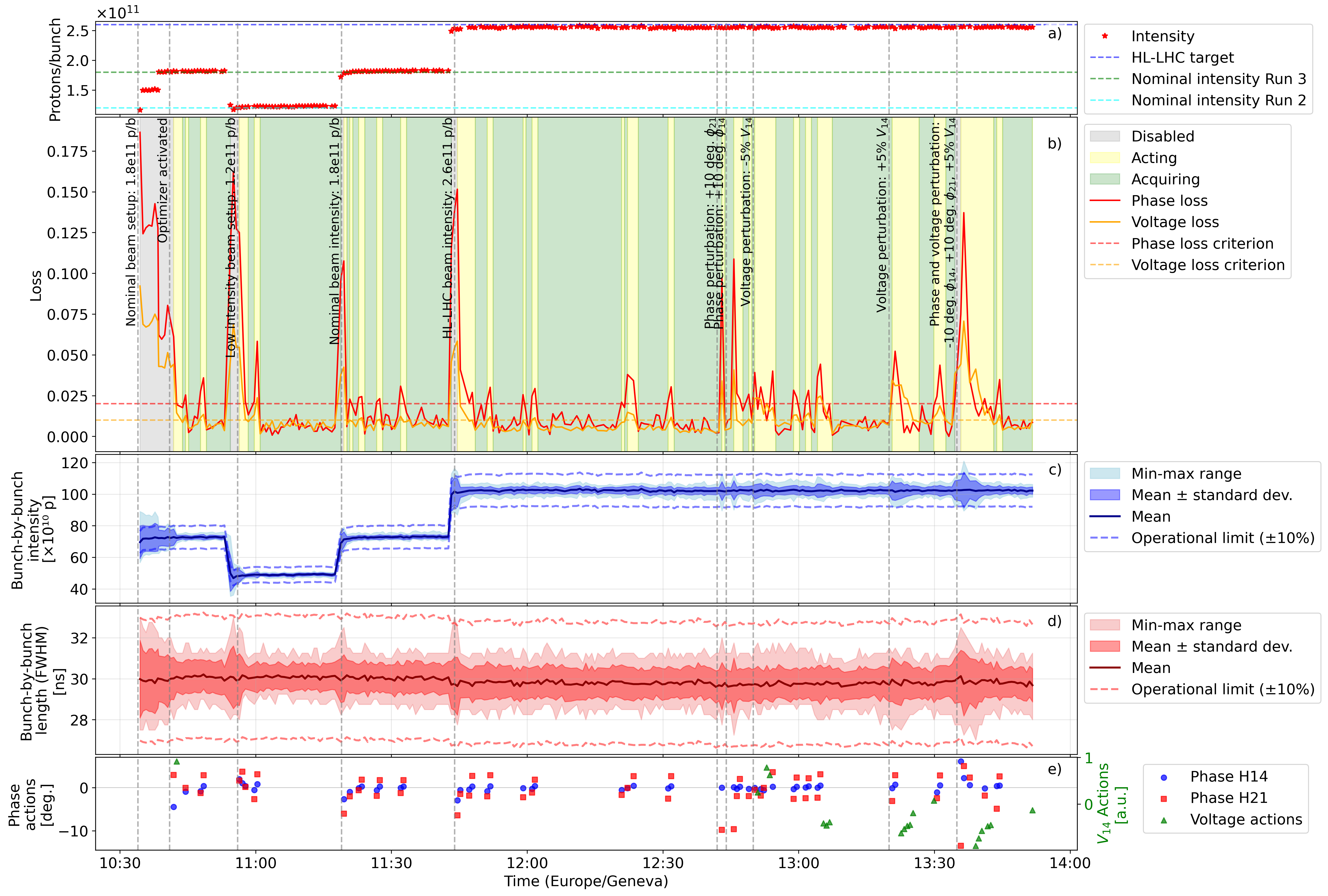}
    \caption{Stress test performed with operational autonomous implementation. From top to bottom, all plots share a common timeline: a) Beam intensity (for this particular beam type, 72 bunches are produced in $h$=84 after triple and double splittings). b) Observed objective function values, $L_{Q}'$ (voltage loss) and $L_{Q-\mathrm{phase}}'$ (phase loss). Dashed horizontal lines indicate the criteria for the losses to be optimized, and vertical horizontal lines mark each action taken known to require re-optimization of the triple splitting. The operational state of the controller is shown as shaded regions: the \textit{acquiring} state in green, signaling good beam quality. The \textit{acting} state in yellow, whenever beam is non-optimal and the controller is requesting a settings change. The \textit{disabled} state is shown in gray, when the controller is not online. c) Bunch-by-bunch intensity spread. d) Bunch-by-bunch length spread. e) Actions requested by the agent for the parameters of the triple splitting: the phase actions in degrees, and the voltage in a normalized unit relative to a reference voltage program. Overall, the plots show that for each perturbing action made, the criteria for the optimizer are pushed above their activation thresholds. The agents then quickly act to restore beam quality within a few steps of interaction, with no need for manual intervention.}
    \label{fig:md_stress_test_results_plot}
\end{figure*}

The system is designed to act only if either of the set criteria are not met, to avoid unnecessarily trimming hardware settings when beam quality is already acceptable. Initial deployment featured the same models and criteria as in the on-demand script. Over the course of operational testing, various minor improvements to the implementation of observations and calculation of criterion were made, although the general architecture of the solution remained unchanged. The initial guess is provided by the feature extractor, followed by iterative updates from the RL agents until the beam reaches target losses. The criteria applied to determine phase and voltage thresholds were adapted from the $L_{Q-\mathrm{phase}}$ and $L_{Q}$ to a term directly derived from the final bunch intensities and lengths:

\begin{gather}
    L_{Q}' = (\frac{|\tau_1| + |\tau_2| + |\tau_3|}{3} +\frac{|I_{1}| + |I_{2}| + |I_{3}|}{3}) /2 \\
    L_{Q-\mathrm{phase}}' = (\frac{|\tau_1| + |\tau_3|}{2} + \frac{|I_{1}| + |I_{3}|}{2})/2
\end{gather}

Here, $I_n$ is the relative bunch intensity and $\tau_n$ is the relative bunch length of bunch $n$, see Eq. (\ref{eq:phase_obs},~\ref{eq:volt_obs}). There were two main motivations for this change. Firstly, to make the loss computation less sensitive to alignment errors between the different bunch profiles. Secondly, to more closely relate the criteria to well-established features like bunch intensity/length variation, rather than the less interpretable values of the $L_{Q-\mathrm{phase}}$ and $L_{Q}$. Furthermore, it was observed that the RL agent optimizing voltage was occasionally converging to a sub-optimal state. This was suspected to be caused by the measured bunch lengths not aligning well with the simulation in the close-to optimal regime for certain beam intensities. This led to a re-training of the voltage agent with only bunch intensities as input.

The first deployment of the fully autonomous, controller-like version of the scheme was completed in early 2025. After an initial test run the model was fully operational including extensive logging of data, and has been available for optimization since.

\section{Results}
\label{sec:results}

\begin{figure*}[!ht]
    \centering
    \includegraphics[width=1.0\linewidth]{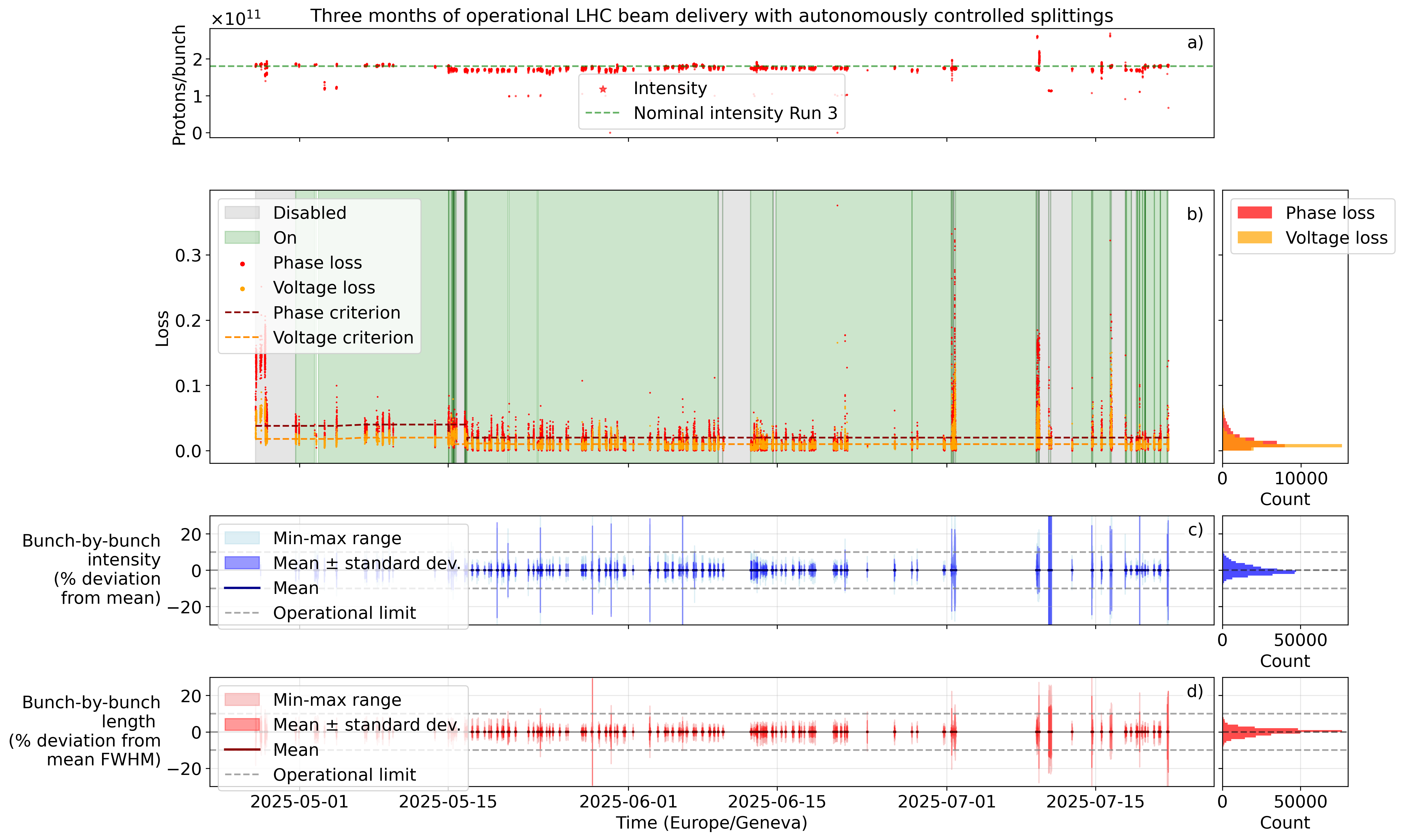}
    \caption{Three months of operational data delivering beam for the LHC after deployment of the fully autonomous scheme. From top to bottom, all plots share a common timeline: a) Beam intensity (for this particular beam, either 48 or 36 bunch schemes were used). b) Observed objective function values, $L_{Q}'$ (voltage loss) and $L_{Q-\mathrm{phase}}'$ (phase loss). Dashed horizontal lines indicate the criteria for the losses to be optimized, and were available to operations for finetuning; note the reduction in thresholds with time. The shaded regions show when the controller was online in green, and disabled in grey. In the c) and d) plots, bunch-by-bunch intensity and length spreads are displayed, as \% deviation from the mean. The histograms on the right are projections, clearly showing a majority of measurements featuring good beam quality. Objective values are low, and small bunch-by-bunch variations, well-within specifications are consistently observed.}
    \label{fig:long_form_data}
\end{figure*}
The original on-demand optimization script was tested under varying beam-conditions validating consistent and fast convergence~\cite{internal-note-jwulff}. To further test the capability of the fully autonomous implementation, a controlled stress test was performed with a nominal 72 bunch LHC-type beam. After an initial period of setting-up beam parameters, the optimizer was enabled. Several challenging beam changes known to require re-optimization of the triple splitting parameters were made by, e.g.,

\begin{itemize}
    \item changing beam intensity,
    \item inducing manual errors in phases, $\phi_{14}$ and $\phi_{21}$,
    \item inducing manual errors in voltage, $V_{14}$.
\end{itemize}

During the test no manual actions were taken to optimize any settings. The controller was allowed to react autonomously to the beam quality degradations as they occurred. A detailed plot summarizing the results is displayed in Fig.~\ref{fig:md_stress_test_results_plot}.

The figure illustrates that for each perturbing action made, the criteria are pushed above their activation thresholds, and the controller autonomously acts to restore beam quality within a few steps of interaction. On both the bunch-by-bunch intensity and length spread plots, a strong correlation between optimizer losses and beam observables is evident.

In addition to the stress test, the long-term performance is demonstrated best by an overview of the logged data from running on the main operational beam to the LHC during several months (Fig.~\ref{fig:long_form_data}).

Whenever the beam quality degrades, fast actions bring the beam quality back into good regions within less than ten iterations, often in a single step. Overall, for the three-month period, low optimizer losses correlate with low spreads in intensity and bunch lengths, indicating good controller performance. 

\section{Summary and conclusions}

In this report we have presented an optimization solution for longitudinal beam splittings, in particlar the triple splitting performed in the CERN PS for LHC-type beams. It is based on a combination of supervised and reinforcement learning methods. Training was performed entirely on simulated data, with domain randomization used to bridge the gap to measurements, and the resulting models were transferred directly to the accelerator. The system was deployed operationally as an on-demand optimizer in 2023, refined through 2024 to handle more challenging beam generation schemes. In 2025 it was extended into a fully autonomous controller that monitors the splitting quality and activates optimization without manual intervention.

Several factors were essential to this success. Decomposing the optimization into separate phase and voltage sub-problems enabled faster development of each RL agent, enabling exploration of observation and reward definitions. It also minimized the risk of unnecessary trims of hardware settings when only one type of error (phase or voltage) was present. Environment hyperparameters such as maximum episode length, reward shape, and termination criteria proved more important to tune than the SAC algorithm hyperparameters themselves. Access to a fast surrogate model derived from the pre-simulated BLonD data made RL training practically possible. The operational deployment of the autonomous controller was only achieved thanks to the flexibility of the new LBO acquisition system, which provided the continuous measurements needed to construct meaningful state represenations on a beam-by-beam basis.

While the system is effective for the current use case, possible extensions remain. The models are trained with a static dataset, generated from a simulation with fixed voltage programs. Therefore, they are inherently biased towards the current operational scheme and any substantial change would require re-simulating the dataset and re-training. More broadly, this maintenance burden of handling re-training under changing conditions is a general drawback of learning-based methods compared to simpler controllers. From an operational point of view, it may be preferable to search for a less complex solution, agnostic to the overall settings of the accelerator: with further feature engineering enabled by the LBO, it may be possible to decouple the optimization parameters sufficiently to allow classical control methods. On the ML-side, promising directions include integrating measured data into training, exploring richer state descriptions, and combining the CNN feature extractor and RL policy into an end-to-end model. None of these were required for the present use case, but each could improve efficiency or generality.

Despite these caveats, the deployed implementation is most effective. It has run on LHC-type beams since 2023, effectively removing the need for manual optimization of the triple splitting, and has led to autonomously maintained, consistent splitting quality across varying beam conditions since 2025. This makes the system a concrete demonstration that RL and ML methods can be used to produce production ready control in an operating accelerator environment, and marks a significant step towards the fully autonomous setup of LHC-type beams.

\begin{acknowledgements}

The authors wish to extend our gratitude and appreciation for the great support of the PS operators, without whom this project would not have been possible. Special thanks is extended to Heiko Damerau and Jake Flowerdew, 
for their advice and comments on the contents and structure of the paper. Further thanks for their contributions at CERN enabling the results presented in this paper is extended to Amaury Beeckman, Alexander Huschauer, Michael Schenk, Verena Kain, Michi Hostettler, Georges Trad, and Adrian Menor. 
Finally, a thanks to all members of the SY-RF-BR section at CERN for their support and discussions. 

\end{acknowledgements}

\bibliography{apssamp}

\begin{thebibliography}{24}
\bibitem [{\citenamefont {Wulff}\ and\ \citenamefont {Lasheen}(shed)}]{internal-note-jwulff}%
  \BibitemOpen
  \bibfield  {author} {\bibinfo {author} {\bibfnamefont {J.~A.}\ \bibnamefont {Wulff}}\ and\ \bibinfo {author} {\bibfnamefont {A.}~\bibnamefont {Lasheen}},\ }\bibfield  {title} {\bibinfo {title} {On-demand optimization of {LHC} beams in the {Proton Synchrotron} using {Reinforcement Learning}},\ }\href@noop {} {\bibfield  {journal} {\bibinfo  {journal} {CERN Technical note}\ } (\bibinfo {year} {To be published})}\BibitemShut {NoStop}%
\end{thebibliography}

\end{document}